\documentclass[amsmath,showpacs,aps,prb,twocolumn]{revtex4}
\usepackage{bm}
\usepackage{graphics}
\usepackage{graphicx}
\pdfoutput=1

\newcommand{\bq}{\begin{equation}}
\newcommand{\ee}{\end{equation}}

\begin{document}

\title{Equilibrium currents in chiral systems with non-zero Chern number}

\author{E.~G.~Mishchenko}
\affiliation{Department of Physics and Astronomy, University of Utah, Salt Lake
City, Utah 84112, USA}

\author{O.~A.~Starykh}
\affiliation{Department of Physics and Astronomy, University of Utah, Salt Lake
City, Utah 84112, USA}

\begin{abstract}
We describe simple quantum-mechanical approach to calculating equilibrium particle current along the edge of a system with 
non-trivial band spectrum topology. The approach does not require any {\em a priori} knowledge of the band topology and, as a matter of fact,
treats topological and non-topological contributions to the edge currents on the same footing. We illustrate its usefulness by demonstrating 
the existence of `topologically non-trivial' particle currents along the edges of three different physical systems: 
two-dimensional electron gas with  spin-orbit coupling and Zeeman magnetic field, surface state of a topological insulator, and 
kagom\'e antiferromagnet with Dzyaloshinskii-Moriya interaction. We describe relation of our results to the notion of orbital magnetization.
\end{abstract}


\maketitle

\section{Introduction}

Orbital contributions to the magnetization in systems with topologically non-trivial band spectrum represent a relatively new but active field of study. The non-local nature of the corresponding quantum operator is one of the obstacles in calculating orbital magnetization of Bloch electrons. This difficulty has been tackled by means of the Wannier representation \cite{XSN,TCV,XYF,CTV}, standard perturbation theory \cite{SVX}, first principles calculation \cite{LVT,CGS}, Keldysh formalism \cite{ZYF}.

In the present paper we develop a different approach based on the equation of motion for the density matrix. We begin by noting that the non-locality of the magnetization ${\bf M}$ is intimately related to the presence of a boundary in the system. In an ininite homogeneous system magnetization would be undefined. It acquires a concrete physical meaning only by virtue of its spatial variation ${\bf M}({\bf r})$ near a boundary or any other inhomogeneity, where it relates to the density of the {\it uncompensated} electric currents\cite{LL8}, ${\bf j} = c\nabla \times {\bf M}$. In other words, the concept of magnetization is simply a different way to represent {\it local} electric currents. The latter, however, can be calculated directly from microscopic theories where their definition does not raise the issues of non-locality at all. Below we use this approach to find equilibrium currents in a number of systems that are characterized by a non-zero Chern number: two-dimensional electron gas with  spin-orbit coupling and Zeeman terms, surface of a topological insulator, and 
kagom\'e antiferromagnet with Dzyaloshinskii-Moriya interaction.

In systems with a significant  spin-orbit splitting in the band structure the spin degree of freedom is tied to the momentum of the particle. As a result, acceleration of the particle leads to non-adiabatic spin precession, which in turn affects the particle's motion (current). This phenomenon was first described by Karplus and Luttinger \cite{KL},
see Refs.~\onlinecite{haldane2004,burkov2014} for a modern perspective,
 in terms of the geometric (Berry) phase  ${\bm \Omega}({\bf k})$ that produces the anomalous velocity $e{\bf E}\times {\bm \Omega}$, responsible for the anomalous Hall effect.
The electric field ${\bf E}$ can, in principle, exist even in equilibrium, for example due to a confining potential of the boundary of a system. In materials with a properly designed non-trivial geometric phase this can lead to the existence of the equilibrium boundary currents. Similar currents could circulate inside the system around defects or impurities. In the present paper we  study the conditions for the occurrence of such currents near a boundary of a two-dimensional electron gas (2DEG) with Bychkov-Rashba spin-orbit interaction, as well as the distribution of current density. Other types of chiral systems are then considered with the same method.

\section{2DEG with Rashba spin-orbit interaction}
\label{sec:rashba}

Let us consider a single-particle Hamiltonian that describes the motion of electrons in a potential $U({\bf r})$
\begin{equation}
\label{ham} H = -\frac{\hbar^2}{2m}  {\bm \nabla}^2 -i \hbar \lambda ~\hat{\bm \eta}\cdot
{\bm \nabla}-h\hat \sigma_z+U({\bf r}),
\end{equation}
in the presence of both the spin-orbit coupling $\lambda$ and Zeeman field $h$.
The matrices $\hat{\bm \eta}
= {\bf z} \times \hat{\bm \sigma}$ are related to the spin Pauli matrices $\hat{\bm \sigma}$, the direction ${\bf z}$ is perpendicular to the plane of 2DEG, and $m$ is the electron the effective mass.  In case where the Zeeman term originates from the coupling of electron spin to the perpendicular magnetic field, $H_z$, the Zeeman field is $h= \frac{eg}{2m_0c} H_z$, where $g$ is the g-factor. In what follows we neglect the effect of the magnetic field $H_z$ on the orbital motion of electrons. Such approximation is justified, for example, when the g-factor is large.  Another realization of this situation is provided by a system of neutral cold atoms where the orbital coupling $(e/c) {\bf p} \cdot {\bf A}$ is absent,
while  the Zeeman interaction is still present.
\begin{figure}[h]
\begin{center}
\includegraphics[scale=0.55]{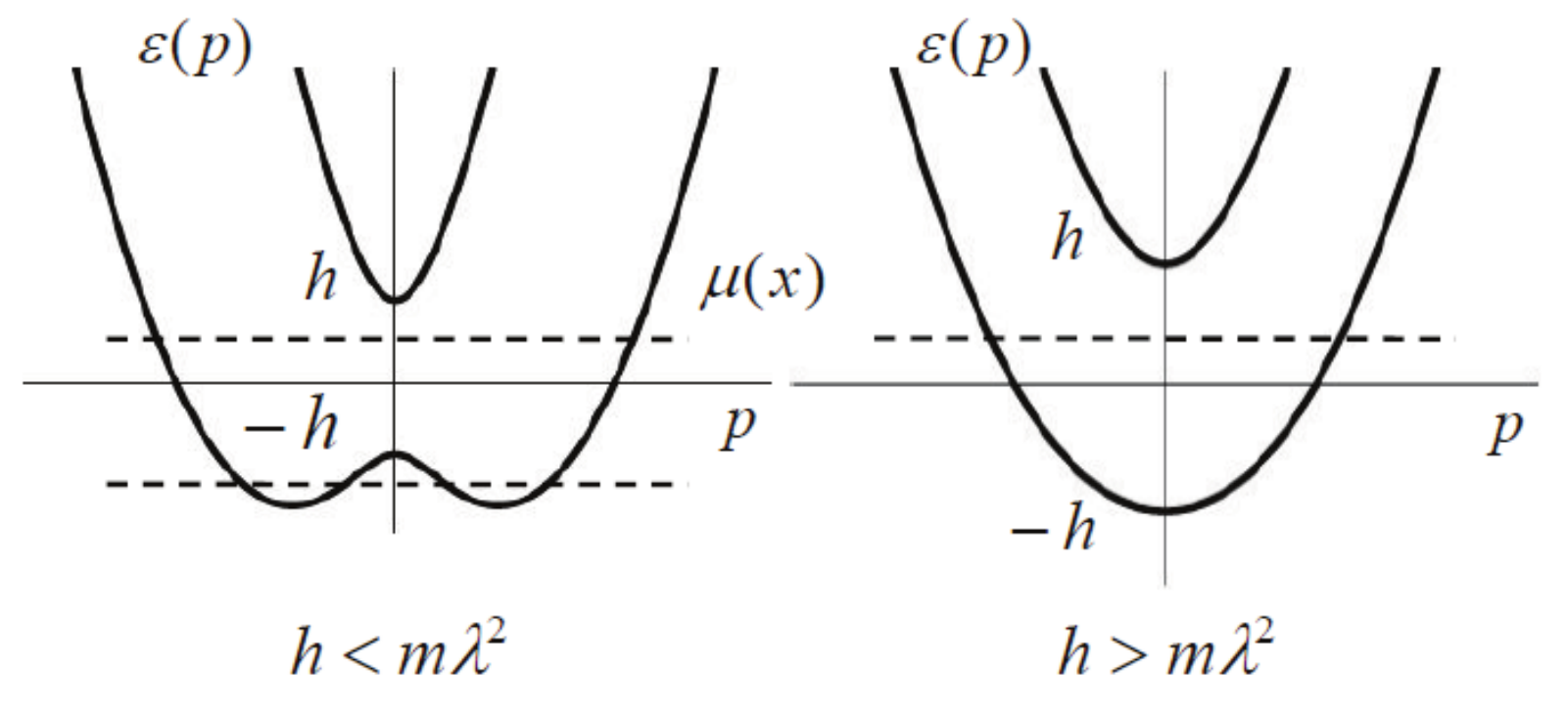}
\caption{Spectrum of a two-dimensional electron gas with spin-orbit interaction and Zeeman splitting, Eq.~(\ref{eq:spectrum}). The left panel shows the spectrum in case of spin-orbit coupling exceeding the Zeeman field:  a ring of minima is formed at a finite value of momentum and a local maximum appears at $p=0$. The right panel corresponds to the case of strong Zeeman coupling: both electron subbands are monotonic functions of momentum. Different possibilities for the position of the chemical potential $\mu(x)$ are indicated by the dashed lines.}
  \label{fig:spectrum}
\end{center}
\end{figure}

From the equation of motion for the electron operators, $\partial \hat \psi/\partial t =i[H, \hat \psi]$, the equation for the density matrix
\begin{equation}
\label{densitym} f_{\alpha \beta}({\bf r},{\bf r'};t) = \langle
 \psi_\beta^\dagger ({\bf r'},t) \psi_\alpha ({\bf r},t)
 \rangle
\end{equation}
can be easily obtained. It is most conveniently written in the
 Wigner representation,
\begin{equation}
\label{wign} \hat f_{\bf p}({\bf R}, t)= \int d {\bm \rho}~ e^{-i {\bf
p} {\bm \rho}} \hat f\left({\bf R}+\frac{\bm \rho}{2}, {\bf R}-\frac{\bm \rho}{2};t \right).
\end{equation}
After straightforward calculation we obtain from Eq.~(\ref{ham}),
\begin{eqnarray}
\label{boltz-rewritten} \frac{\partial
\hat f_{\bf p}}{\partial t}+\frac{1}{2} \left\{  \frac{\bf p}{m} +
\lambda \hat {\bm \eta}, \nabla \hat f_{\bf p} \right\} +i\lambda p
[\hat{\eta}_{\bf p},\hat f_{\bf p }]\nonumber \\ -ih[\hat\sigma_z,\hat f_{\bf p}] +i\int d{\bf q}~U_{\bf q} (\hat
f_{{\bf p}-\frac{\bf q}{2}}-\hat f_{{\bf p}+\frac{\bf
q}{2}})e^{i{\bf qR}}=0
\end{eqnarray}
where  $\hat \eta_{\bf p}=\hat{\bm \eta}\cdot {\bf n}_p$ is the
projection of the spin operator $\hat{\bm \eta}$ onto the direction
of the electron momentum ${\bf n}_p$.
In case when the typical distance over which the potential $U({\bf R})$ changes smoothly 
(the implied condition is discussed at the end of the present Section), the last term in Eq.~(\ref{boltz-rewritten}) can
be cast in a more familiar spatial gradient form,
\begin{equation}
\label{boltz-rewritten1} \frac{1}{2} \left\{  \frac{\bf p}{m} +
\lambda \hat {\bm \eta}, \nabla \hat f_{\bf p} \right\} +i\lambda p
[\hat{\eta}_{\bf p},\hat f_{\bf p }]-ih[\hat\sigma_z,\hat f_{\bf p}]-{\bm \nabla} U \cdot \frac{\partial
\hat f_{\bf p}}{\partial {\bf p}}=0.
\end{equation}
As we are interested in currents in a steady state (equilibrium),  the time derivative has been dropped in the last equation.

The smooth potential $U ({\bf R})$ determines the position of the bottom of
the band in the vicinity of the system's edge. Correspondingly, in the zeroth order in the gradient ${\bm \nabla} U$  the density matrix is
given by its equilibrium form
\begin{equation}
\label{equilibrium}
\hat f_{\bf p}^{(0)}=\frac{1}{2}(f_+ + f_-) +\frac{1}{2}(f_+-f_-)\frac{\lambda p ~\hat \eta_{\bf p} - h \hat \sigma_z}{\sqrt{\lambda^2p^2 +h^2}},
\end{equation}
where the Fermi-Dirac distributions for the two subbands are
\begin{equation}
\label{FD}
f_\pm=\frac{1}{\exp{[\frac{\varepsilon_\pm (p)+U({\bf R})-\zeta}{T}]}+1}.
\end{equation}
The two spin-split subbands,
\begin{equation}
\varepsilon_\pm (p)=\frac{{ p}^2}{2m} \pm \sqrt{\lambda^2 { p}^2 +h^2}
\label{eq:spectrum}
\end{equation}
are non-degenerate at ${\bf p}=0$ due to the effect of the Zeeman field, see Fig.~\ref{fig:spectrum}. Note that $\zeta$ is the {\it electrochemical} potential, which is  constant throughout the whole system. At zero temperature it indicates where the filled states are separated  from the empty states with respect to their {\it total} energy. It is also useful to consider the position-dependent {\it chemical} potential,  $\mu({\bf R})=\zeta -U({\bf R})$, which separates filled and empty states with respect to the ``kinetic'' energy (total energy sans the potential energy of the edge). In particular, the chemical potential $\mu({\bf R})$ is more convenient when the distribution of momenta is needed (as opposed to the distribution of the total energies for which the electrochemical potential $\zeta$ is a more natural choice).

To obtain the non-adiabatic correction to the distribution function, linear in ${\bm \nabla} U$, we write,
$\hat f_{\bf p}=\hat f_{\bf p}^{(0)}+\hat f_{\bf p}^{(1)}$, and neglect gradients of the correction, $\hat f_{\bf p}^{(1)}$, keeping the latter only in the ``precession'' terms:
\begin{equation}
\label{linearized}
[\lambda p\hat{\eta}_{\bf p}-h\hat\sigma_z,\hat f^{(1)}_{\bf p }] =i{\cal
K}_{\bf p},
\end{equation}
where the right-hand side contains the gradients of $\hat f_{\bf p}^{(0)}$ and $U({\bf R})$,
\begin{equation}
\label{K}
\hat{\cal K}_{\bf p}=\frac{1}{2} \left\{  \frac{\bf p}{m} +
\lambda \hat {\bm \eta}, {\bm \nabla} \hat f_{\bf p}^{(0)} \right\} -{\bm \nabla} U \cdot \frac{\partial
\hat f_{\bf p}^{(0)}}{\partial {\bf p}}.
\end{equation}
The solution of Eq.~(\ref{linearized}) is readily found in the matrix form:
\begin{equation}
\label{zeroq} \hat f_{\bf p}^{(1)}= \frac{i(\lambda p~\hat{\eta}_{\bf p}-h\hat\sigma_z)\hat{\cal K}_{\bf p}}{2(\lambda^2 p^2+h^2)}.
\end{equation}
Substituting now the adiabatic approximation (\ref{equilibrium}) into Eq.~(\ref{K}) and then into Eq.~(\ref{zeroq}), after simple but somewhat lengthy algebra, we arrive at the gradient correction,
\begin{eqnarray}
\label{f1}
\hat f_{\bf p}^{(1)}=  -\frac{ \lambda {\bm \nabla} U \cdot [ \lambda p ({\bf n}_p\times {\bm
\sigma})+h\hat {\bm \sigma} ]}{4(\lambda^2 p^2+h^2)^{3/2}} \Bigl[f_+-f_-  \nonumber\\
-(f'_++f'_-) \sqrt{\lambda^2 p^2+h^2}  \Bigr],
\label{eq:f1}
\end{eqnarray}
where the notation $f'$ stands for the derivative of the Fermi-Dirac distribution with respect to its energy argument.

We are now ready to evaluate the electric current propagating along the edge of the system. It consists of two terms,
\begin{equation}
\label{current_two}
{\bf j} =  e\text{Tr}\sum_{\bf p} \left(\frac{\bf p}{m} +\lambda {\bm \eta}\right)\hat f_{\bf p}-\frac{eg}{4m_0}\hat{\bf z} \times \nabla \text{Tr}\sum_{\bf p} \hat{\sigma}_z \hat f_{\bf p}.
\end{equation}
The first, orbital, term, originating from the electron velocity $\partial H / \partial {\bf p}$, is determined by the correction (\ref{f1}) to the distribution function.
The second, paramagnetic, term describes the current produced by the inhomogeneous  distribution of spin density and is determined, as calculated below,
by the equilibrium distribution function \eqref{equilibrium}.

Below we consider the two contributions to the current (\ref{current_two}) separately. Our main result is that {\it each} contribution vanishes when electrons are present in both the upper and lower subbands, $h<\mu({\bf R})$, but  are {\it nonzero} when 
only the lowestsubband is populated, $\mu({\bf R}) < h$, so that $f_+=0$.

\begin{figure}[h]
\begin{center}
\includegraphics[scale=0.65]{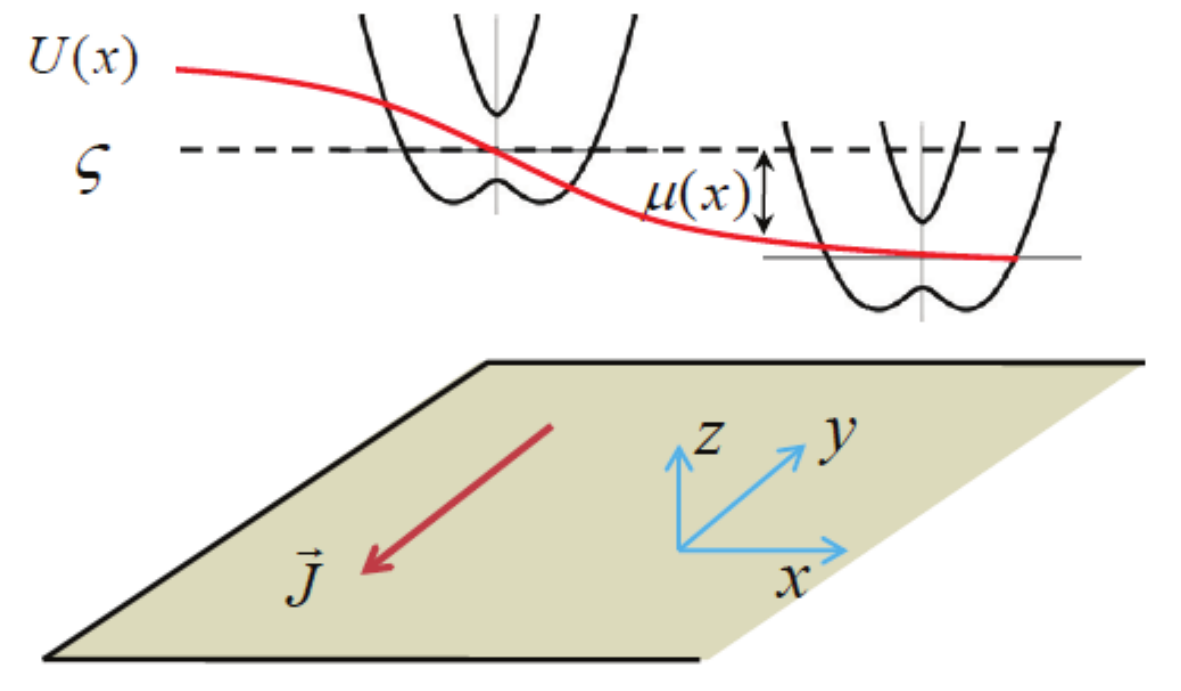}
\caption{Smooth boundary of a two-dimensional electron gas. The upper panel illustrates the depletion of the electron density near the edge. The dashed line indicates the position of the electrochemical potential $\zeta$ as counted from the bottom of the band deep inside 2DEG. The chemical potential $\mu(x)$ is a function of the coordinate ($\mu(\infty)=\zeta$). The lower panel indicates the direction of the equilibrium electric edge currents ${\vec J}$. }
  \label{fig:edge}
\end{center}
\end{figure}

We choose the boundary to coincide with the ${\bf y}$ axis of the system so that the $x$-coordinate measures a distance from the edge, see Fig.~\ref{fig:edge}.
The potential energy near the edge creates electric field $e{\bf E}_{\rm edge} = - {\bf x} (\partial U/\partial x)$ which is ultimately responsible for the equilibrium current flowing
along the boundary of the sample.

The orbital contribution to the current along the edge is
 $j^{(1)}_y=e \text{Tr}\sum_{\bf p}(p_y/m+\lambda \hat \sigma_x) \hat f_{\bf p}^{(1)}$.
Using Eq.~(\ref{f1}), taking the trace and carrying out the angular integration we obtain
 \begin{eqnarray}
\label{eq:j1}
j^{(1)}_y(x) = - \frac{e}{4\pi}\lambda^2 h \frac{\partial U}{\partial x} \int\limits_0^\infty
\frac{pdp}{(\lambda^2 p^2+h^2)^{3/2}} \nonumber\\ \times
\left( f_+-f_-- (f'_++f'_-) \sqrt{\lambda^2 p^2+h^2} \right).
\end{eqnarray}
This simple expression contains very rich physics: its integrand is determined by the
Berry curvature,
\begin{equation}
\label{eq:Omega}
{\bm \Omega}_{\beta}({\bf p}) =  {\bf z}  \frac{\beta\lambda^2 h}{(\lambda^2 {p}^2+h^2)^{3/2}}
\end{equation}
opposite for the lower, $\beta = -1$, and upper, $\beta = +1$,
subbands. The curvature (\ref{eq:Omega}) is non-zero only when both the spin-orbit and Zeeman splittings are present. Note that the non-trivial band topology appears naturally in our calculations rather than being assumed to exist.

After simple integration (see Appendix~\ref{app:integration} for details) we find
\begin{equation}
\label{current_low}
j^{(1)}_y(x)=\frac{e}{4\pi} \frac{\partial U}{\partial x} \left\{ \begin{array}{cc} 0,~~~ & h<\mu(x),\\
1-h/{\cal H}(\mu),~~~ & -h<\mu(x)< h, \\
-2h/{\cal H}(\mu),~~~ & \mu(x)< -h, \end{array} \right.
\end{equation}
where ${\cal H}(\mu) =\sqrt{h^2+m^2\lambda^4+2m\lambda^2 \mu(x)}$.

The second contribution to the current in Eq.~(\ref{current_two}) is due to the inhomogeneous spin density and appears already in the adiabatic approximation,
when $\hat f_{\bf p}$ is replaced with $\hat f_{\bf p}^{(0)}$, Eq.~(\ref{equilibrium}). It can be written as
${\bf j}^{(2)}({\bf r}) = c {\bm \nabla} \times {\bf M}_{\rm para}$, where {\em paramagnetic} magnetization ${\bf M}_{\rm para} = M_{\rm para} {\bf z}$
has the standard form,
\begin{eqnarray}
\label{Mpara}
&&M_{\rm para} = \frac{g\mu_B}{2} \text{Tr} \sum_{\bf p} \hat{\sigma}_z \hat f_{\bf p}^{(0)} = \nonumber\\
&& = -\frac{g e}{2 m_0 c} \int \frac{d^2{ p}}{(2\pi)^2} \frac{h (f_+ - f_-)}{\sqrt{\lambda^2 { p}^2+h^2}} .
\end{eqnarray}
This part is distinct from (\ref{current_low}) in that it is proportional to the extra $g$-factor (in addition to the one implicit in the Zeeman field $h$). Simple calculation gives
\begin{equation}
\label{current_spin}
j^{(2)}_y(x)=\frac{eg}{8\pi}\frac{m}{m_0} \frac{\partial U}{\partial x} \left\{ \begin{array}{cc} 0,~~~ & h<\mu(x),\\
h/{\cal H}(\mu),~~~ & -h<\mu(x)< h, \\
2h/{\cal H}(\mu),~~~ & \mu(x)< -h. \end{array} \right.
\end{equation}

The applicability of the gradient approximation (\ref{boltz-rewritten1}) to the exact equation (\ref{boltz-rewritten}) for the density matrix requires that the relevant Fermi components of the boundary potential are smooth on the scale of the electron wavelength, $q\ll p_F$, taken at the Fermi level. If the width of the edge  is  $L_{edge}$ this condition implies that
\begin{equation}
\label{condition1}
p_F L_{edge}  \gg 1.
\end{equation}
Thus  the chemical potential should not be too close to the bottom of the band, where the Fermi momentum $p_F$ vanishes.

The second condition arises from our use of the expansion in powers of the gradient of the potential energy $\nabla U$, Eqs.~(\ref{equilibrium}) and (\ref{f1}). Each subsequent term in this expansion acquires an extra power of ${ \lambda { \nabla} U}/{(\lambda^2 p^2+h^2)}$. Since the non-zero net current is found when only one subband is occupied, the typical momenta of interest are $p \sim m\lambda^2$, and the required condition can be written in the form,
\begin{equation}
\label{condition2}
\lambda { \nabla} U \ll \text{max} (m^2 \lambda^4,h^2).
\end{equation}
Note that the two conditions (\ref{condition1}) are (\ref{condition2}) are essentially the same for the most interesting situation where the chemical potential in the bulk of the 2DEG is inside the Zeeman gap and $h\sim m\lambda^2$. The width of the edge $L_{edge}$ is the distance over which the density of electrons changes from its bulk value to zero. In that case, $U\sim  m \lambda^2 \sim h$, the Fermi momentum, $p_F \sim m \lambda$, and both conditions coincide.

\subsection{Net current}

It is now easy to calculate the net current, $J_y=\int_{-\infty}^\infty j_y (x) dx$, propagating along the edge. Since ${\partial U}/{\partial x}=-{d\mu }/dx$, the net current is expressed in terms of the chemical potential deep inside the system, which also conicides with the electrochemical potential  $\zeta\equiv \mu(\infty)$, when the boundary potential is chosen to vanish there, $U(\infty)=0$. Integration of Eq.~(\ref{current_low}) yields, for various possible values of $\zeta$,
\begin{equation}
\label{current_net}
J^{(1)}_y=\frac{e}{4\pi} \left\{ \begin{array}{cc} 0,~~~ & h<\zeta,\\
\frac{h[{\cal H}(\zeta)-h]}{m\lambda^2}-\zeta,~~~ & -h<\zeta< h, \\
2h\frac{{\cal H}(\zeta)}{m\lambda^2},~~~ & \zeta< -h, \end{array} \right.
\end{equation}
Similarly, the net current due to the inhomogeneous spin density is
\begin{equation}
\label{current_net_spin}
J^{(2)}_y=-\frac{egh}{8\pi}\frac{m}{m_0} \left\{ \begin{array}{cc} 0, & h<\zeta,\\
({\cal H}(\zeta)+m\lambda^2-h)/(m\lambda^2), & -h<\zeta< h, \\
2{\cal H}(\zeta)/(m\lambda^2), & \zeta< -h. \end{array} \right.
\end{equation}

Note that the form of the spectrum depends on whether the spin-orbit energy $m\lambda^2$ is greater or smaller than the Zeeman energy $h$, see Fig.~\ref{fig:spectrum}.
If the Zeeman energy is the larger of the two there is never a situation where the lower subband has the region of the negative group velocity and, as a consequence,
two Fermi circles. If this is the case the range $\mu(x)<-h$ is absent.
The equations (\ref{current_low})-(\ref{current_net_spin}) are still applicable in this case as long as  the expressions for $-h<\mu(x)$ are used.

\section{Topological insulators}

The formalism of the preceding Section can be applied to other two-dimensional systems with chiral Hamiltonians that are linear in momentum, such as graphene or the surface of a topological insulator. Due to its sublattice symmetry and the ensuing presence of the two Dirac points with the opposite Berry curvatures the net currents tend to vanish in graphene. However, since on the surfaces of topological insulators (TI) such Dirac points are also spatially separated, the currents are non-zero  \cite{pesin}.

The spectrum of the 2D electron gas on the surface of TI in the perpendicular magnetic field is still given by Eqs.~(\ref{ham}) and (\ref{eq:spectrum}) where the formal limit of $m \to \infty$ should be taken. The spin-orbital coupling $\lambda$ now acquires the meaning of the Fermi velocity. The smooth potential $U(x)$ can be produced by means of electrostatic gates placed above the surface (since in the Dirac approximation the electron band is ``bottomless'', the true boundary, or edge, can not be envisioned).

Using now the general expression Eq.~(\ref{eq:j1}) we quickly arrive at the conclusion that the current vanishes unless the chemical potential lies inside the Zeeman gap, $-h < \mu (x) < h$, in which case only the term $ f_-$ in the integrand contributes to the current:
\begin{equation}
\label{TIorbital}
j^{(1)}_y(x) = - \frac{e}{4\pi} \frac{\partial U}{\partial x} \Theta(h-|\mu(x)|).
\end{equation}
This result has previously been derived in Ref.~\onlinecite{pesin}.
When the potential drop is large enough so that a $p$-$n$ junction is created, with the Fermi level residing above the gap on one side and below it on the other side of the junction, the net current $eh/2\pi$ is flowing along the junction.
\begin{figure}[h]
\begin{center}
\includegraphics[scale=0.46]{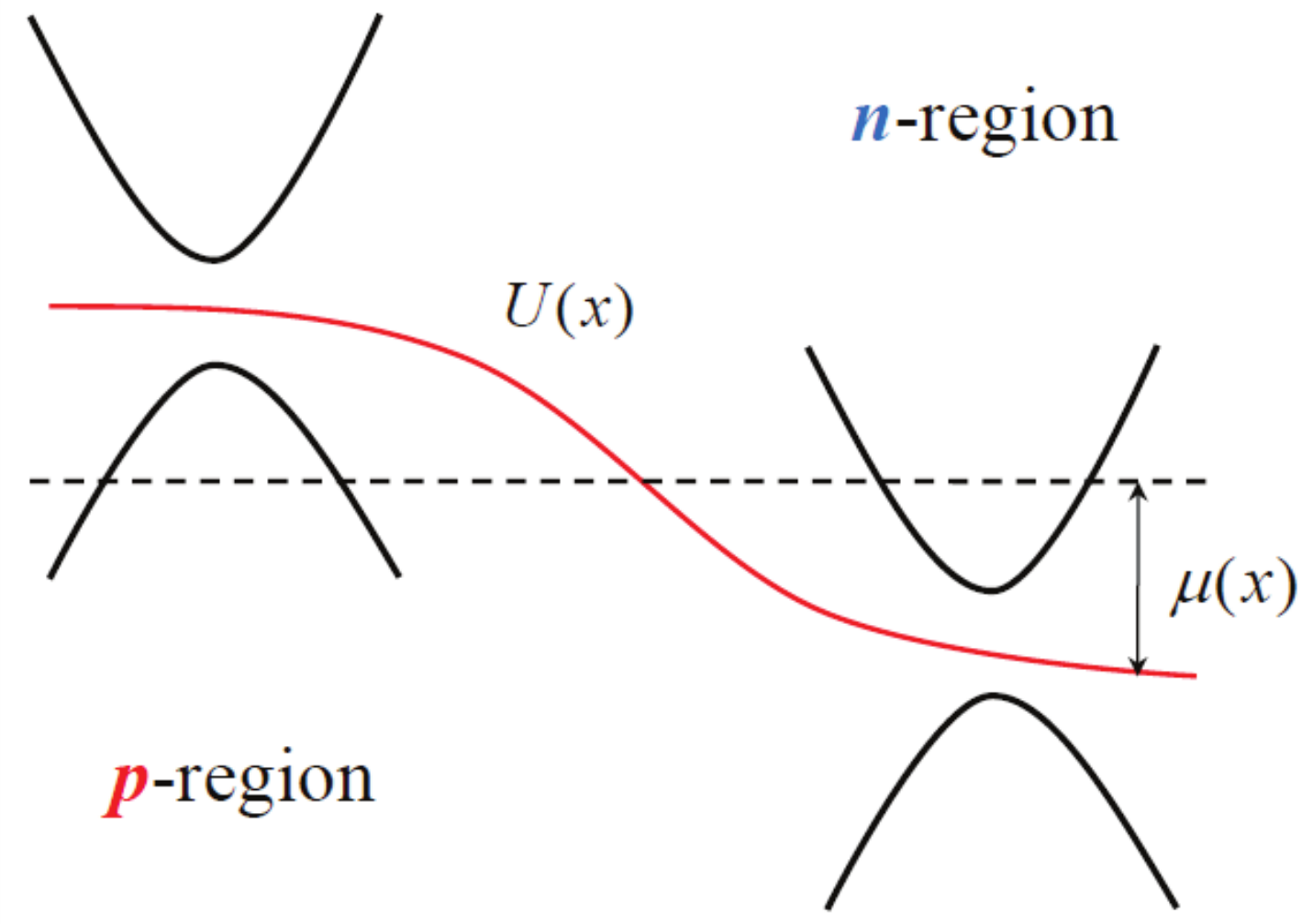}
\caption{Two-dimensional electron gas formed by the surface states of a topological insulator. Smooth gate potential creates a $p$-$n$ junction. The net current $eh/2\pi$ is flowing along the junction. }
  \label{fig:TI}
\end{center}
\end{figure}

We now turn to the second, paramagnetic, contribution to the current, Eq.~(\ref{current_general}). While in case of conventional 2DEG it is in general of the same order as the orbital term, cf.~Eqs.~(\ref{current_low}) and (\ref{current_spin}), in TI the paramagnetic term is significantly smaller. Nonetheless, this contribution is important since it has a completely different dependence on the chemical potential. In particular, it is nonzero where the orbital contribution (\ref{TIorbital}) vanishes. Calculating the spatial derivative of the magnetization (\ref{Mpara}) we observe that only the vicinity of the Fermi surface contributes to the momentum integral via the derivatives of the equilibrium distribution functions $f_\pm$. As a result we obtain,
\begin{equation}
\label{TIspin}
j^{(2)}_y(x) = \frac{egh}{8\pi m_0 \lambda} \frac{\partial U}{\partial x} \left\{ \begin{array}{cc} 1,~~~ & h<\mu(x),\\
0,~~~ & -h<\mu(x)< h, \\
-1,~~~ & \mu(x)< -h. \end{array} \right.
\end{equation}
In particular the paramagnetic current is of opposite sign in the $p$ and $n$ regions of the $p$-$n$ junction. This should be contrasted with the orbital part (\ref{TIorbital}), which is nonzero only within the ``neutral'' strip of the junction.

\section{Kagom\'e antiferromagnet with Dzyaloshinskii-Moriya interaction above the saturation field}

\begin{figure}[h]
\begin{center}
\includegraphics[scale=0.35]{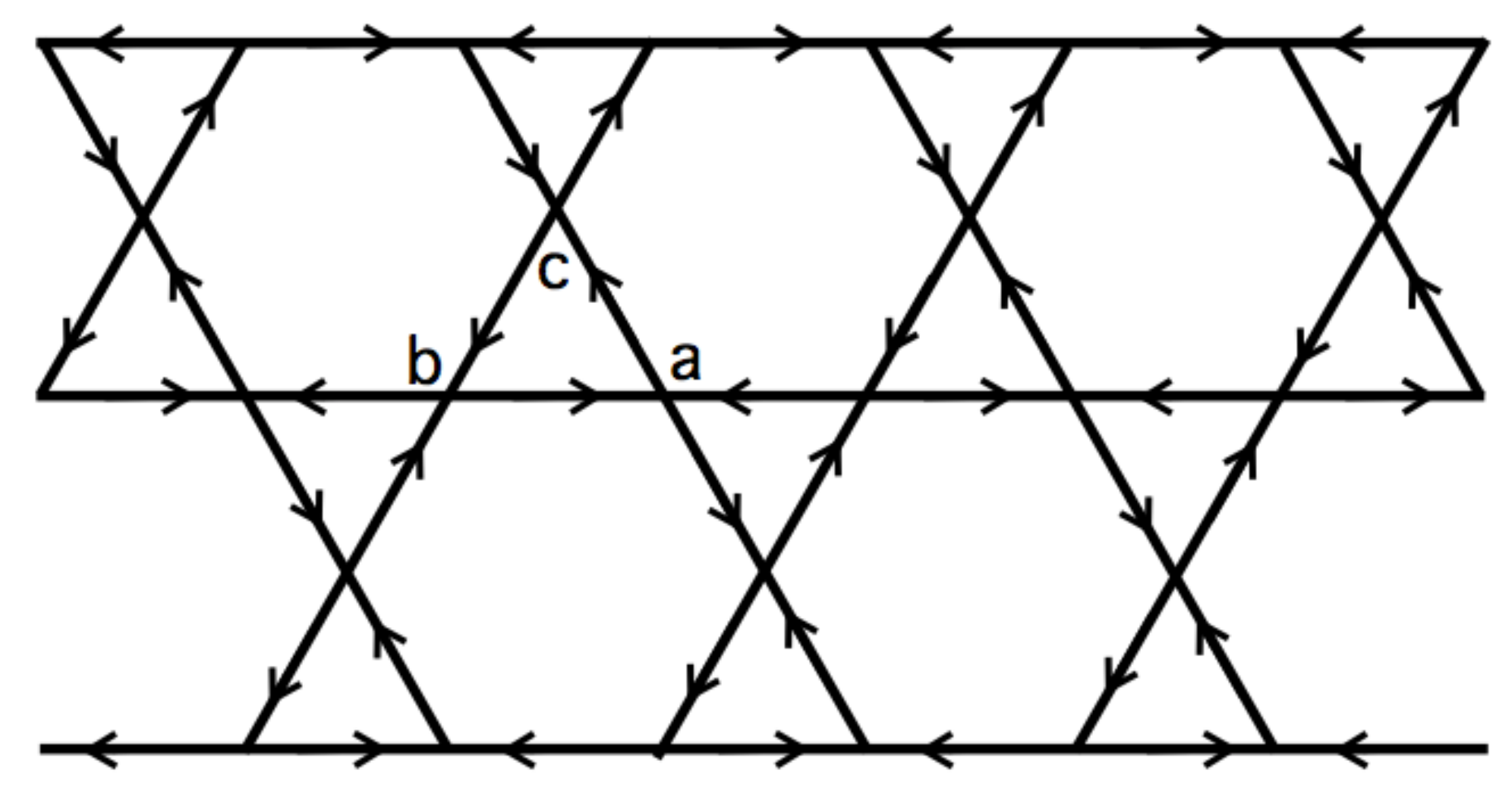}
\caption{Kagom\'e lattice antiferromagnet. Bond arrows point from site $i$ to site $j$ in DM interaction term $D {\hat z}\cdot {\bf S}_i \times {\bf S}_j$.
Also indicated are sublattices $a$, $b$ and $c$.}
  \label{fig:kagome}
\end{center}
\end{figure}

The edge current does not need to be that of electrons only. Here we show that a very similar physics plays out in a rather different system - a
two-dimensional insulating quantum antiferromagnet on kagom\'e lattice  in the presence of  external magnetic field. The current that flows around the edge in this
case is that of charge-less magnons, which are quanta of excitations of the angular momentum, i.e. spin waves. The role of spin-orbit interaction is played
by the Dzyaloshinskii-Moriya (DM) interaction ${\bf D}_{ij}\cdot {\bf S}_i \times {\bf S}_j$, where spatial vector ${\bf D}_{ij}$ is living on the bond $(ij)$
connecting the nearest neighbor sites of the kagom\'e lattice. We choose DM vectors ${\bf D}_{ij} = D \hat{z}$ to be normal to the layer,
and oriented along the bonds $(ij)$ of the kagom\'e lattice as shown in Fig.~\ref{fig:kagome}.
Note that this choice respects translational and rotational $C_6$ (rotations about the center of the hexagon) symmetries of the lattice,
and is of the kind realized in ZnCu$_3$OH$_6$Cl$_2$ \cite{mila2009,mei2011}. Closely related to it kagom\'e ferromagnet system
is currently under investigation \cite{chisnell2014}.
Similar set-ups, in relation to thermal Hall effect, have
been recently discussed in \cite{onose,murakami}.

We subject kagom\'e antiferromagnet to a strong magnetic field ${\bf B} = B \hat{z}$ which exceeds the saturation field $B_{\rm sat}$ above which the
spins are fully polarized. Excitations of this fully polarized ground state are spin waves which we describe with the help
of a standard large-S approximation
\begin{equation}
S^z_{\bf r} = S - a^\dag_{\bf r} a_{\bf r} , S^\dag \approx \sqrt{2S} a_{\bf r} .
\end{equation}
Since the unit cell of kagom\'e lattice contains 3 spins, there are in fact three kinds of spin waves, one for each sublattice type, which
we denote as $a_{\bf r}, b_{\bf r}$ and $c_{\bf r}$ in the following. The coordinate ${\bf r}$ here is that of the unit cell.
Simple algebra shows that the linear spin wave Hamiltonian of the system has a $3\times3$ matrix form
\begin{equation}
\label{eq:kag0}
H_{\rm kagome} = 2JS \sum_{\bf k} (a^\dag_{\bf k}, b^\dag_{\bf k}, c^\dag_{\bf k}) {\cal M}_{\bf k}
\left(\begin{array}{c}
a_{\bf k}\\
b_{\bf k}\\
c_{\bf k}
\end{array}\right)
\end{equation}
where the matrix reads
\begin{equation}
{\cal M}_{\bf k} = \left(
\begin{array}{ccc}
h-2 & (1+i \tilde{d}) \cos\frac{k_1}{2} & (1-i \tilde{d}) \cos\frac{k_2}{2} \\
(1-i \tilde{d}) \cos\frac{k_1}{2}& h-2 & (1+i \tilde{d}) \cos\frac{k_3}{2}  \\
(1+i \tilde{d}) \cos\frac{k_2}{2} & (1-i \tilde{d}) \cos\frac{k_3}{2} & h-2 \\
\end{array} \right)
\end{equation}
and $k_1 = 2 k_x, k_2 = k_x + \sqrt{3} k_y, k_3 = k_x - \sqrt{3} k_y$. Here $\tilde{d} = D/J$ is dimensionless DM interaction and $h = g\mu_B B/(2 J S)$
is rescaled magnetic field.

This Hamiltonian possesses remarkable extensive degeneracy in the absence of DM interaction ($d=0$) -- its lowest energy band is completely
flat, $\epsilon_1({\bf k}) = h - 3$. \cite{mzh2007} Finite DM, $d\neq 0$, lifts the degeneracy and provides $\epsilon_1$ with a weak dispersion, see \eqref{eq:kag1} below.
One of the eigenmodes of the Hamiltonian \eqref{eq:kag0} is a symmetric precession mode with high energy of the order of $h_{\rm sat} = 3$, while the two others,
describing relative fluctuations of spins on different sublattices, have much smaller energy of the order $h - h_{\rm sat} \ll h_{\rm sat}$.
As a result, near the saturation ($h \geq h_{\rm sat}$) and at low temperature $T$, one can just project the high-energy precession out.
Carrying this approximation out and focusing on the long wavelength limit ${\bf k} \to 0$ leads us to a much simple $2\times2$ effective Hamiltonian
\begin{equation}
\tilde{H} = \sum_k (\psi_{1,{\bf k}}^\dag, \psi_{2,{\bf k}}^\dag) \tilde{\cal M}_{\bf k}
\left(\begin{array}{c}
\psi_{1,{\bf k}}\\
\psi_{1,{\bf k}}
\end{array}\right)
\end{equation}
where
\begin{equation}
\tilde{\cal M}_{\bf k} = \left(
\begin{array}{cc}
V + \frac{k_y^2}{m} & -\frac{k_x k_y}{m} - i \sqrt{3}d \\
-\frac{k_x k_y}{m} + i\sqrt{3} d & V + \frac{k_x^2}{m}
\end{array} \right)
\end{equation}
Here we denoted $V = 2 J S (h - h_{\rm sat})$, $m = 8/(2JS)$ and $d = 2J S \tilde{d}$. Eigenvalues of this Hamiltonian are
\begin{equation}
\epsilon_{\pm}({\bf k}) = V + \frac{{\bf k}^2}{2 m} \pm \sqrt{3d^2 + \frac{{\bf k}^4}{4m^2}}
\label{eq:kag1}
\end{equation}
Observe that for $d=0$ the lowest eigenvalue becomes flat, $\epsilon_{-} \to V$, in accordance with the discussion above.

Subsequent analysis is based on Eq.~\eqref{eq:kag1} and, as we show, valid at low temperatures such that $T \ll d$, when
thermal occupation of $\beta = \pm$ bands with energies $\epsilon_{\pm}({\bf k})$ is exponentially small. To describe the
boundary, which we again assume to run along the $y$-axis, we promote $V$ to a position-dependent variable $V\to V({\bf R})$
which smoothly increases from its minimum $V_0 = h - h_{\rm sat}$ value in the bulk of the magnet to $V\to\infty$ on the vacuum side, ${\bf R} = (X,Y) \to (\infty,Y)$.
Long but straightforward algebra leads to the bosonic analogue of Eq.~\eqref{boltz-rewritten1}
\begin{eqnarray}
&&\{ \frac{i (p_y \hat{\sigma}_1 + p_x \hat{\sigma}_3)}{m}, \frac{\partial f_{\bf k}}{\partial X}\} +
\frac{\partial V}{\partial X} \frac{i \partial f_{\bf k}}{\partial k_x} - \frac{i k_x}{m} \frac{\partial f_{\bf k}}{\partial X} \nonumber\\
&& + [\sqrt{3}d\hat{\sigma}_2 - \frac{k_x k_y}{m}\hat{\sigma}_1 + \frac{k_y^2 - k_x^2}{2m} \hat{\sigma}_3, f_{\bf k}] = 0
\end{eqnarray}
Neglecting linear gradients $\partial /\partial X$ at first, we find $f_{\bf k} \to \hat f_{\bf k}^{(0)}$ where
\begin{equation}
\label{equilibrium1}
\hat f_{\bf k}^{(0)}=\frac{1}{2}(f_+ + f_-) -\frac{1}{2}(f_+-f_-)\frac{\vec{t}_{\bf k} \cdot \vec{\hat{\sigma}}}{\sqrt{3 d^2 + ({\bf k}^2/2m)^2}},
\end{equation}
and $\vec{t}_{\bf k} = (k_x k_y/m, -\sqrt{3}d, (k_x^2 - k_y^2)/(2m))$. The equilibrium distribution function is now Bose-Einstein,
\begin{equation}
f_{\beta=\pm}({\bf k}) = (\exp[\epsilon_{\beta}({\bf k})/T] +1)^{-1}.
\end{equation}
Note that $V$ plays the role of chemical potential now.

The correction is found to be
\begin{eqnarray}
\hat f_{\bf k}^{(1)}&=&-\frac{\partial V}{\partial X} \frac{f_+ - f_- - (f'_+ +f'_-) |t_{\bf k}|}{4|t_{\bf k}|^3} \nonumber\\
&&\times\Big(\frac{k_y {\bf k}^2}{2 m^2} \hat{\sigma}_2 +
\frac{\sqrt{3}d}{m} (k_x \hat{\sigma}_1 - k_y \hat{\sigma}_3)\Big)
\end{eqnarray}
The velocity along the boundary is $v_y = \partial \tilde{\cal M}_{\bf k}/\partial k_y$ and the magnon current density is then
$j_y = \sum_{\bf k} {\text{Tr}}(v_y \hat{f}_{\bf k})$,
\begin{eqnarray}
\label{eq:magnon-current}
j_y &=& \frac{\partial V}{\partial X}\sum_{\bf k} \frac{\sqrt{3}d {\bf k}^2}{2m^2[3d^2 + ({\bf k}^2/2m)^2]^{3/2}}\\
&&\times \Big(f_+ - f_- - (f'_+ +f'_-)\sqrt{3d^2 + ({\bf k}^2/2m)^2}\Big) \nonumber
 \end{eqnarray}
Once again, the kernel of this expression is given by the Chern curvature of the two magnon bands involved.

We now focus on the low-temperature regime, $T \ll d$, when the temperature is much smaller than the splitting
between the magnon bands. In this case only the lowest, $\beta=-$, band needs to be retained in \eqref{eq:magnon-current}.
Focusing on the total magnon current, and denoting $z = k^2/(2m)$, we find
\begin{eqnarray}
I_y &=& - \frac{\sqrt{3} d}{2\pi} \int_0^\infty dX \frac{\partial V}{\partial X} \int_0^\infty \frac{d z ~z}{(3 d^2 + z^2)^{3/2}} \nonumber\\
&&\times[f_-(z) + f'_-(z) \sqrt{3d^2 + z^2}] .
\end{eqnarray}
The upper limit of the $z$-integration can be set to infinity due to the exponential convergence of the integral in the $T \ll d$ limit.
Simple calculation shows that under these conditions the second term in square brackets dominates, and we find
\begin{equation}
I_y = \frac{\sqrt{3} d}{2\pi} \left(\frac{T}{\sqrt{3} d}\right)^2 e^{-(V_0 - d)/T}.
\end{equation}
The contribution from the first term is smaller by additional factor of $T/d \ll 1$.

\section{Discussion}
\label{sec:disc}

The fact that the currents found are {\it equilibrium} and non-dissipative (in all cases considered ${\bf j} \cdot {\bm \nabla} U = 0$) makes
one want to ask, what kind of magnetization ${\bf M}$ such a current gives rise to? The standard steady-state relation
${\bf j}({\bf r}) = c {\bm \nabla} \times {\bf M}$ implies that ${\bf M}$ is not exhausted by ${\bf M}_{\rm para}$, expressed by Eq.~\eqref{Mpara}.
One can expect that there is an additional {\em orbital} magnetization ${\bf M}_{\rm orb}$, which is responsible for the contribution ${\bf j}^{(1)}$ given by Eq.~\eqref{current_low}, via a similar relation,
${\bf j}^{(1)} = c {\bm \nabla} \times {\bf M}_{\rm orb}$.
This additional part of the magnetization was initially introduced on the basis  of the semi-classical wave packet considerations in Refs.~\onlinecite{XSN,TCV} and
later derived rigorously in Ref.~\onlinecite{SVX}. To complement our density matrix calculations we present a detailed application of that formalism to the Rashba system of Sec.~\ref{sec:rashba} in Appendix~\ref{app:orb}.

We emphasize that our main result, however,  is not the application of the standard relation ${\bf j}({\bf r}) = c {\bm \nabla} \times {\bf M}$
to the particular cases of systems with the non-trivial band topology. Rather, our findings point to a novel way to experimentally observe
``topological'' contribution (orbital magnetization) ${\bf M}_{\rm orb}$ and to separate it, via the difference in the $g$-factor dependence of the currents ${\bf j}^{(1,2)}$,
from the standard paramagnetic magnetization ${\bf M}_{\rm para}$. The experimental technique of this kind has recently been developed \cite{nowack2013,nowack2014}.

Another important application of our calculations is the system of cold atoms, where the resulting edge current represents a {\em mass current}, circulating around the boundary of the system, which should be observable \cite{gurarie}.
Such mass current is determined by ${\bm \nabla} U$, which is routinely controlled in cold atom systems. This leads to the realistic possibility of studying
current generation in response to a change in the confining potential $U({\bf R})$ and/or Zeeman potentials.
It can also be detected by a muon spin rotation experiments, like in Sr$_2$RuO$_4$ \cite{luke1998}.

As we have shown, a circulating edge current of magnons is also realized in a kagom\'e antiferromagnet geometry, which too can now
be realized in optical lattices \cite{jo2012}. Perhaps more importantly, our calculation raises an intriguing possibility of generating
circulating magnon currents around a non-magnetic impurity.

\begin{acknowledgements}
We would like to thank Dima Pesin for numerous insightful discussions of the magnetization current and surface states of topological insulators
and Oleg Tchernyshyov for discussions of the kagom\'e antiferromagnet and its non-trivial band structure in the presence of DM interactions,
and pointing out $\mu$SR paper \cite{luke1998}. We also thank K. Nowack, V. Gurarie and M. Hermele for useful discussions.
This work is supported by the National Science Foundation through grant DMR-12-06774 (O.A.S.) and by the Department of Energy,
Office of Basic Energy Sciences, Grant No.~DE-FG02-06ER46313 (E.G.M).

\end{acknowledgements}

\appendix
\section{Calculation of the electric current}
\label{app:integration}


{\it Velocity contribution}.
Substituting Eq.~(\ref{f1}) into the expression for the electric current, $j_y=e \text{Tr}\sum_{\bf p}(p_y/m+\lambda \hat \sigma_x) \hat f_{\bf p}^{(1)}$, and calculating the trace and the angle integral we arrive at the remaining integral over the absolute value of the electron momentum,
\begin{eqnarray}
\label{current_general}
j_y(x) = - \frac{e}{4\pi}\lambda^2 h \frac{\partial U}{\partial x} \int\limits_0^\infty
\frac{pdp}{(\lambda^2 p^2+h^2)^{3/2}} \nonumber\\ \times
\left( f_+-f_-- (f'_++f'_-) \sqrt{\lambda^2 p^2+h^2} \right).
\end{eqnarray}

i) When the chemical potential lies above the bottom of the upper subband,  $\mu(x)>h$, the contribution from the difference of the two Fermi-Dirac functions in Eq.~(\ref{current_general}) at zero temperature is
\begin{equation}
\label{I1}
I_1=\int\limits_0^\infty
pdp \frac{f_+-f_-}{(\lambda^2 p^2+h^2)^{3/2}}= -\frac{1}{\lambda^2} \left(\frac{1}{\omega_+}-\frac{1}{\omega_-}\right),
\end{equation}
where $\omega_\pm =\sqrt{\lambda^2 p_\pm^2 +h^2}$ and $p_\pm$ are the Fermi momenta of the upper and lower subbands determined from the equations, $\varepsilon_\pm (p) = \mu$.
Similarly, the contributions from the derivatives of the Fermi-Dirac functions (at $T=0$ given simply by delta-functions) in Eq.~(\ref{current_general}) are,
\begin{equation}
\label{I2}
I_2=- \int\limits_0^\infty
{pdp}\frac{f'_++f'_-}{\lambda^2 p^2+h^2}=\sum_\pm \frac{m}{\omega_\pm (\omega_\pm \pm m\lambda^2)}.
\end{equation}
The sum of the two contributions is thus,
\begin{equation}
I_1+I_2= \frac{1}{\lambda^2} \left( \frac{1}{\omega_--m\lambda^2}- \frac{1}{\omega_++m\lambda^2}\right).
\end{equation}
From the condition $p_\pm^2/2m\pm \sqrt{\lambda^2 p_\pm^2+h^2}=\mu (x)$ we find that
\begin{equation}
\omega_\pm=\sqrt{h^2+m^2\lambda^4+2m\lambda^2 \mu(x)} \mp m\lambda^2,
\end{equation}
so that $I_1+I_2=0$, which means that the current density vanishes when $\mu(x)>h$.

ii) When the chemical potential resides inside the Zeeman gap, $-h<\mu(x)<h$, the upper band is completely empty, $f_+=0$, so that only the lower subband contribution should be retained in the expression (\ref{I2}) for $I_2$. In the other integral (\ref{I1}) a similar procedure yields:
 \begin{equation}
I_1=-\int\limits_0^\infty
pdp \frac{f_-}{(\lambda^2 p^2+h^2)^{3/2}}= -\frac{1}{\lambda^2} \left(\frac{1}{h}-\frac{1}{\omega_-}\right).
\end{equation}
The integral in Eq.~(\ref{current_general}) is therefore given by,
 \begin{equation}
I_1+I_2= \frac{1}{\lambda^2} \left( \frac{1}{\omega_--m\lambda^2}- \frac{1}{h}\right),
\end{equation}
giving the current density Eq.~(\ref{current_low}).

iii) Finally when the chemical potential is below the Zeeman gap, $\mu(x)<-h$, the lower subband is occupied only for the momenta in the range $p_1<p<p_2$, whose boundaries are determined by the roots of the equation, $p^2/2m -\sqrt{\lambda^2 p^2+h^2}=\mu (x)$. Similarly to Eq.~(\ref{I1}) we obtain
\begin{equation}
\label{I1low}
I_1=-\int\limits_{p_1}^{p_2}
 \frac{pdp}{(\lambda^2 p^2+h^2)^{3/2}}= -\frac{1}{\lambda^2} \left(\frac{1}{\omega_1}-\frac{1}{\omega_2}\right),
\end{equation}
where
\begin{equation}
\omega_{1,2}=m\lambda^2\mp \sqrt{h^2+m^2\lambda^4+2m\lambda^2 \mu(x)}.
\end{equation}
The second  term Eq.~(\ref{I2}) has now two contributions from the two Fermi momenta $p_1$ and $p_2$ of the lower subband:
\begin{equation}
\label{I2low}
I_2= \frac{m}{\omega_1 (m\lambda^2-\omega_1)}+\frac{m}{\omega_2 (\omega_2 - m\lambda^2)}.
\end{equation}
As a result we obtain,
 \begin{equation}
I_1+I_2= \frac{2}{\lambda^2 \sqrt{h^2+m^2\lambda^4+2m\lambda^2 \mu(x)}},
\end{equation}
reproducing the last line of Eq.~(\ref{current_low}).

{\it Inhomogeneous spin density contribution}. The integral in the expression for the current, Eq.~(\ref{current_two}), with the distribution function given by Eq.~(\ref{equilibrium}), is particularly simple. In case when both subbands are populated,
\begin{equation}
\text{Tr}\sum_{\bf p} \sigma_z \hat f_{\bf p}=-\frac{h (\omega_--\omega_+)}{2\pi \lambda^2}.
\end{equation}
When the local Fermi level is in the Zeeman gap $\omega_+$ has to be replaced with $0$; when the Fermi level is below the gap, $\mu(x) < -h$, we have to replace $\omega_+$ with $\omega_1$ and $\omega_-$ with $\omega_2$. As a result we find Eq.~(\ref{current_net_spin}) which, upon formally replacing $\zeta$ with $\mu(x)$ and differentiating over $x$ yields Eq.~(\ref{current_spin}).

\section{Orbital magnetization}
\label{app:orb}

Applied to the Rashba system of Section~\ref{sec:rashba}, orbital magnetization ${\bf M}_{\rm orb} = M_{\rm orb} {\bf z}$ reads
\begin{equation}
M_{\rm orb} = \sum_{\beta = \pm} \int \frac{d^2 {\bf p}}{(2\pi)^2} \Big( m_\beta f_\beta + \frac{e}{\hbar} \Omega_\beta (\mu - \epsilon_\beta)f_\beta \Big).
\end{equation}
Here $m_\beta$ is the orbital moment of the sub band $\beta$
\begin{equation}
m_\beta = \frac{i e}{2\hbar} \langle \frac{\partial u_\beta}{\partial {\bf p}}| \times (\epsilon_\beta - H_{\bf p}) |  \frac{\partial u_\beta}{\partial {\bf p}}\rangle,
\end{equation}
where $\times$ stands for vector product and $u_\beta$ is the periodic part of the Bloch wave function of the sub band $\beta$,
\begin{eqnarray}
|u_\beta \rangle &=& \frac{1}{\sqrt{2}} \frac{1}{\sqrt{\Delta_{\bf p}^2 + \beta h \Delta_{\bf p}}}
  \left(
\begin{array}{c}
\lambda( p_y + i p_x)\\
h +  \beta \Delta_{\bf p} \\
\end{array}
\right),
\end{eqnarray}
where we abbreviated $\Delta_{\bf p} = \sqrt{h^2 + \lambda^2 {\bf p}^2}$.
Here $H_{\bf p}$ is the Hamiltonian acting on $u_\beta$, so that
$ H_{\bf p} - \epsilon_\beta = \lambda(p_y \hat{\sigma}_x - p_x \hat{\sigma}_y) - h \hat{\sigma}_z - \beta \Delta_{\bf p} \hat{\sigma}_0.$
Simple calculation shows that $m_\beta = e \lambda^2 h/(2(\lambda^2 p^2 + h^2))$ is in fact $\beta$-independent.

Using this and Eq.~\eqref{eq:Omega}  we obtain explicit form
\begin{eqnarray}
M_{\rm orb} &=& \int \frac{d^2 {\bf p}}{(2\pi)^2} \Big\{ \frac{e \lambda^2 h}{2 \Delta_{\bf p}^2} (f_- + f_+) -
\frac{ e \lambda^2 h}{2 \Delta_{\bf p}^3} (\mu - \epsilon_-({\bf p})) f_- \nonumber\\
&& + \frac{ e \lambda^2 h}{2 \Delta_{\bf p}^3} (\mu - \epsilon_+({\bf p})) f_+ \Big\} .
\end{eqnarray}

It is now a simple exercise to check that
\begin{equation}
j_y^{(1)} = - c \frac{\partial M_{\rm orb}}{\partial x}  = -c \frac{\partial M_{\rm orb}}{\partial \mu} \frac{\partial \mu}{\partial x}
\end{equation}
gives {\em exactly} the current density \eqref{eq:j1}. In doing so it is important to remember that there $f'$  stands
for the derivative of the distribution function with respect to its energy argument, and thus
$f'_\pm = \partial f_\pm/\partial \epsilon_\pm = - \partial f_\pm/\partial \mu$.

\end{document}